\documentclass[aps,prd,preprint,superscriptaddress,amsmath,amssymb,showpacs]{revtex4-1}
\usepackage{dcolumn}
\usepackage{graphicx}
\usepackage{float}
\usepackage{physics}
\usepackage{xcolor}
\usepackage[colorlinks=true,allcolors=blue]{hyperref}

\graphicspath{{figs/}}
\begin{document}


\title{Contribution of Coulomb interaction to elastic $pp$ and $p\bar{p}$ scattering in holographic QCD}


\author{Yu-Peng Zhang}
\affiliation{School of Nuclear Science and Technology, University of South China, Hengyang 421001, China}

\author{Xun Chen}
\email{chenxunhep@qq.com}
\affiliation{School of Nuclear Science and Technology, University of South China, Hengyang 421001, China}

\author{Xiao-Hua Li}
\email{lixiaohuaphysics@126.com}
\affiliation{School of Nuclear Science and Technology, University of South China, Hengyang 421001, China}

\author{Akira Watanabe}
\email{watanabe@usc.edu.cn}
\affiliation{School of Mathematics and Physics, University of South China, Hengyang 421001, China}

\date{\today}

\begin{abstract}
The differential cross sections of elastic proton-proton $(pp)$ and proton-antiproton $(p\Bar{p})$ scattering are studied in a holographic QCD model, considering the strong and Coulomb interaction in the Regge regime. Based on previous studies of strong interactions described in terms of Pomeron and Reggeon exchange, we add the contribution of Coulomb interaction described by photon exchange. We present the momentum transfer dependence of the contribution rates for each component, especially for the Coulomb-nuclear interference, which refers to the cross term between both interactions. For the adjustable parameters for the strong interaction, we can adopt the values determined in previous studies, and there are no extra adjustable parameters that need to be determined for the Coulomb interaction. It is presented that the resulting differential cross sections are consistent with the data for $pp$ and $p\Bar{p}$ scattering.
\end{abstract}



\maketitle

\section{Introduction}
Quantum chromodynamics (QCD) is a well-established theory of the strong interactions, and in principle, all strong-interaction phenomena should be describable in terms of the fundamental QCD Lagrangian. In $pp$ and $p\Bar{p}$ scattering with high center-of-mass energy $s$ and small momentum transfer $t$, some practical difficulties were encountered due to the complexity and nonperturbative nature in the soft kinematic region (also called the forward region)~\cite{in1,in2}. The analytical solution of nonperturbative QCD is a challenging task, and QCD is unable to directly deduce various hadronic properties. Before the establishment of QCD, there is a time-honored theory, the Regge theory, which provided a useful framework to analyze the hadron-hadron scattering cross sections~\cite{in3,in4,in5,in6,in7,in8,in9}. The Regge theory, which incorporates both Reggeon and Pomeron contributions, remains a reliable framework for describing the total cross sections of hadronic scattering. The $pp$ and $p\Bar{p}$ scattering amplitudes were described by the Reggeon trajectories and the soft Pomeron~\cite{in10,in11,in12,in13}. The Regge theory is founded on the complex angular momentum analysis. The $2^{++}$ glueball with mass is recognized to be the lightest state on the leading Pomeron trajectory, which has an intercept slightly greater than 1. The growing trend in total cross sections relative to the center-of-mass energy $\sqrt{s}$ is attributed to Pomeron exchange. On the contrary, the exchange of the Reggeon trajectories accounts for the decreasing behavior. For decades, high energy hadron-hadron scattering has been one of the most important research topics in the high energy physics since its cross sections reflect the internal structure of the involved hadrons.

Holographic QCD, a nonperturbative methodology for QCD, has been established employing the anti-de Sitter/conformal field theory (AdS/CFT) correspondence~\cite{in14,in15,in16}. This correspondence, which establishes a connection between a four-dimensional conformal field theory and a gravitational theory in higher-dimensional AdS space, provides us with a hopeful way to investigate strongly coupled quantum field theories. A holographic QCD model has been utilized to examine the spectrum and configuration of hadrons~\cite{in17,in18,in19,in20,in21,in22,in23,in24,in25,in26,in27,in28,in29}, achieving favorable outcomes. And, this model has also been employed to study high energy scattering processes~\cite{in30,in31,in32,in33,in34,in35,in36,in37,in38,in39,in40,in41,in42}. A holographic QCD model has been proposed, relying on string theory, to portray the experimental data of elastic $pp$ and $p\Bar{p}$ scattering cross sections in the Regge regime~\cite{in43,in44,Reggeized}. Scattering amplitudes of hadrons within the Regge regime can be computed via exchanges involving the lightest mesons or glueballs. Subsequently, the single particle propagators are substituted with their Reggeized counterparts, which are obtained through comparison with string scattering amplitudes~\cite{in43,Reggeized,liu2023pomeron}.

As widely recognized, the electromagnetic effect -- soft photon radiation and Coulomb scattering -- is an indivisible component of any strong interaction involving charged hadrons. The strong interaction is conventionally referred to as \lq\lq nuclear\rq\rq or \lq\lq hadronic,\rq\rq while the electromagnetic interaction is commonly known as the \lq\lq Coulomb" interaction. Sometimes these effects obstruct the detection of strong-interaction phenomena but sometimes they present a distinctive source of information on important details of hadronic amplitudes. The holographic QCD is a theoretical framework that relates a four-dimensional conformal field theory to a gravitational theory in the higher dimensional AdS space. In this theoretical framework, the Coulomb interaction is not a crucial constituent, but it is important when the scattering angle is very small (i.e., $|t|$ is very small). In previous studies, this contribution is of importance at $|t|\approx0.002$ $\text{GeV}^2$ and becomes negligible when $|t|>0.01$ $\text{GeV}^2$~\cite{in45,in46}. Moreover, the combined scattering amplitude receives a third contribution reflecting the cross term with both strong and Coulomb exchanges. This term, along with the intricate nature of the scattering amplitudes, characterizes the influence of Coulomb-nuclear interference (CNI) on the differential cross section. The experimental study of CNI in $pp$ and $p\Bar{p}$ scattering can reveal the amplitude structure of hadrons~\cite{in49,in50,in51,in52,in53,in54,in55,in56,Prochazka:2017tby}. Regrettably, the range of scattering angles where such interference is clearly observable is relatively limited, nevertheless, analysis of the differential cross section of this interval can give us some very important information about the strong interaction.

In this work, we study the elastic $pp$ and $p\Bar{p}$ scattering in the Regge regime, taking into account both the strong and Coulomb interaction. This work is an extension of previous research~\cite{liu2023pomeron}, in which the strong interaction described by the Pomeron and Reggeon exchange was considered. The Pomeron exchange makes a major contribution to the cross section of the high energy region and the Reggeon exchange gives the dominant contribution to cross sections in the lower energy region. However, the Coulomb interaction contribution needs to also be considered to describe the data in a very small momentum transfer $|t|$ interval. In our work, the Coulomb interaction is described by the pure real photon exchange QED amplitude, and being affected by the multiphoton exchange process will result in additional phase difference $\alpha\phi$. The Coulomb phase has been studied by many researchers~\cite{Bethe:1958zz,WYphase,simila1,simila2,Cahn,Kopeliovich,Nurushev}. In the present study the interference formulas of R. Cahn~\cite{Cahn} were adopted for taking into account the CNI effects. We explicitly show how both of the interaction contributions vary with the energy, focusing on the contribution ratios. It is presented that the resulting differential cross sections are consistent with the data in a wide kinematic region for both the $pp$ and $p\Bar{p}$ scattering.

The present paper is structured as follows. The model, focusing on both interactions is outlined in Sec.~\ref{2}. We briefly review the formalism developed in the preceding studies, and present the expressions for the total scattering amplitude and differential cross sections. In Sec.~\ref{3}, the $|t|$ dependence of these contributions is shown in detail, focusing on the contribution ratios. We also show the data analysis of the differential cross sections. Our conclusion with the implications of this work is given in Sec.~\ref{4}.

\section{Model setup}\label{2}
\subsection{Strong interaction amplitude in holographic QCD}
In the previous study~\cite{liu2023pomeron}, the contribution of combining both Pomeron and Reggeon was considered in the Regge regime, which was described by Reggeized spin-2 glueball and vector meson, respectively. The strong-interaction amplitudes can be written in the following form

\begin{equation}
    F_N=F_g+F_{\nu}\label{Nuclear Amplitude}.
\end{equation}

The amplitude of the glueball exchange can be obtained by combining the proton-glueball-proton vertex~\cite{Pagels:1966zza} and the massive spin-2 glueball propagator~\cite{Propagator}, and similarly, the amplitude of the vector meson exchange is obtained by combining the proton-vector-proton vertex and the vector meson propagator~\cite{Reggeized}.

Hence, the strong-interaction amplitude can be written as
\begin{equation}
\begin{aligned}
F_N= & \frac{-i \lambda_{g}^{2}}{8\left(t-m_{g}^{2}\right)}\left[2 s A^{2}(t)\left(\bar{u}_{1} \gamma^{\alpha} u_{3}\right)\left(\bar{u}_{2} \gamma_{\alpha} u_{4}\right)+4 A^{2}(t) p_{2}^{\alpha} p_{1}^{\beta}\left(\bar{u}_{1} \gamma_{\alpha} u_{3}\right)\left(\bar{u}_{2} \gamma_{\beta} u_{4}\right)\right] \\
& +\frac{i \lambda_{v}^{2}}{t-m_{v}^{2}} \eta_{\mu \nu}\left(\bar{u}_{1} \gamma^{\mu} u_{3}\right)\left(\bar{u}_{2} \gamma^{\nu} u_{4}\right),
\end{aligned}
\end{equation}
where $t=-(p_3-p_1)^2$, $\lambda_g$ is the proton-glueball-proton coupling constant, $\lambda_{v}$ is the proton-vector-proton coupling constant, $m_{g}$ is the mass of glueball, and $m_{v}$ is the vector meson mass. At $t\rightarrow0$, the form factors  $A(0)\rightarrow1$.

By taking the modulus and the spin averaged sum of strong-interaction amplitude, the differential cross section has the following form:
\begin{equation}
\begin{aligned}
\frac{d \sigma_N}{d t} & =\frac{1}{16 \pi s^{2}}\left| F_N(s, t)\right|^{2} \\
& =\frac{\lambda_{g}^{4} s^{2} A^{2}(t)}{16 \pi\left|t-m_{g}^{2}\right|^{2}}-\frac{\lambda_{g}^{2} \lambda_{v}^{2} A^{2}(t) s}{4 \pi\left|t-m_{g}^{2}\right|\left|t-m_{v}^{2}\right|}+\frac{\lambda_{v}^{4}}{4 \pi\left|t-m_{v}^{2}\right|^{2}}.
\end{aligned}
\end{equation}
Here, the differential cross section only contains the lightest states, in order to include higher spin states on the Pomeron and Reggeon trajectories, the Reggeized procedures are employed~\cite{Reggeized}. The propagator of the massive spin-2 glueball is to be replaced by
\begin{equation}
    \frac{1}{t-m_g^2}\quad\to\quad\left(\frac{\alpha_g'}{2}\right)e^{-\frac{i\pi\alpha_g(t)}{2}}\frac{\Gamma\left[-{\chi_g}\right]\Gamma\left[1-\frac{\alpha_g(t)}{2}\right]}{\Gamma\left[-{\chi_g}-1+\frac{\alpha_g(t)}{2}\right]}\left(\frac{\alpha_g's}{2}\right)^{\alpha_g(t)-2},
\end{equation}
where $\chi_{g}=2 {\alpha_g^{\prime}}{{m_p}^{2}+\frac{3}{2} \alpha_g(0)}-3 $,
 and the dependence on $\chi_{g}$ is introduced. The propagator of the vector meson is to be replaced by
\begin{equation}
    \dfrac{1}{t-m_v^2}\quad\to\quad\alpha_v'\:e^{-\frac{i\pi\alpha_v(t)}{2}}\:\sin\left[\dfrac{\pi\alpha_v(t)}{2}\right]\:(\alpha_v's)^{\alpha_v(t)-1}\:\Gamma[-\alpha_v(t)]\:.
\end{equation}
The differential cross section of both the Pomeron and Reggeon exchange can be obtained by replacing the factors $\frac{1}{t-m_{v}^{2}}$ and $\frac{1}{t-m_{g}^{2}}$.
 The invariant amplitude for strong interaction can be expressed as
\begin{equation}
\begin{aligned}
F_N(s, t)= & -s \lambda_{g}^{2} A^{2}(t) e^{-\frac{i \pi \alpha_{g}(t)}{2}} \frac{\Gamma\left[-\chi_{g}\right] \Gamma\left[1-\frac{\alpha_{g}(t)}{2}\right]}{\Gamma\left[\frac{\alpha_{g}(t)}{2}-1-\chi_{g}\right]}\left(\frac{\alpha_{g}^{\prime} s}{2}\right)^{\alpha_{g}(t)-1} \\
& +2s \lambda_{v}^{2} \alpha_{v}^{\prime} e^{-\frac{i \pi \alpha_{v}(t)}{2}} \sin \left[\frac{\pi \alpha_{v}(t)}{2}\right]\left(\alpha_{v}^{\prime} s\right)^{\alpha_{v}(t)-1} \Gamma\left[-\alpha_{v}(t)\right] .
\end{aligned}
\end{equation}
In the above equation, there are seven adjustable parameters in total. Three of these parameters are related to the Pomeron exchange, i.e., the intercept $\alpha_g(0)$, slope $\alpha^{\prime}_g$ and proton-glueball coupling constant $\lambda_g$. For these adjustable parameters to the Pomeron exchange, we use the values given in previous work~\cite{xie2019elastic}, $\alpha_g(0)=1.084$, $\alpha_g^\prime=0.368$ $\rm GeV^{-2}$, and $\lambda_g=9.59$ $\rm GeV^{-1}$. For the other four adjustable parameters, the intercept $\alpha_v(0)$, slope $\alpha^{\prime}_v$ , $pp$ scattering coupling constant $\lambda_v^{pp}$, and $p\Bar{p}$ scattering coupling constant $\lambda_v^{p\Bar{p}}$ are taken from Ref.\cite{liu2023pomeron}, $\alpha_v(0)=0.444$, $\alpha_v^\prime=0.9257$ $\rm GeV^{-2}$, $\lambda_v^{pp}=7.742$ $\rm GeV^{-1}$, and $\lambda_v^{p\Bar{p}}=16.127$ $\rm GeV^{-1}$. In the domain of strong interaction, similar to previous investigations~\cite{xie2019elastic}, we employ the proton gravitational form factors for $A(t)$  calculated from the soft-wall model.

\subsection{Electromagnetic form factor}
In this paper we apply the AdS/QCD model to the region of small momentum transfer $0 \leq|t| \leq 0.01$ $\rm GeV^{2}$ in which the contribution of the Coulomb interaction is not negligible. To the Coulomb interaction, we introduce the electromagnetic form factor of proton which was derived from the authors of Ref.~\cite{formfactors} by considering a Dirac field coupled to a vector field in the five-dimensional AdS space in the AdS/QCD model . We use the results obtained from the soft-wall model, in which the AdS geometry is smoothly cut off by a background dilaton field at the infrared boundary. And the final expression for the form factor does not bring any adjustable parameter.
The metric of five-dimensional AdS space is expressed as
\begin{equation}
d s^2=g_{M N} d x^M d x^N=\frac{1}{z^2}\left(\eta_{\mu \nu} d x^\mu d x^\nu-d z^2\right),
\end{equation} 
where $\eta_{\mu \nu}=\operatorname{diag}(1,-1,-1,-1)$, and $\mu, \nu=0,1,2,3$. $z$ is the fifth coordinate ranging from $0$ to $\infty$. The model action is given by
\begin{equation}
\begin{aligned}
S_F= & \int d^{d+1} x \sqrt{g} e^{-\Phi(z)}\left(\frac{i}{2} \bar{\Psi} e_A^N \Gamma^A D_N \Psi\right. \\
& \left.-\frac{i}{2}\left(D_N \Psi\right)^{\dagger} \Gamma^0 e_A^N \Gamma^A \Psi-(M+\Phi(z)) \bar{\Psi} \Psi\right),\label{1}
\end{aligned}
\end{equation}
where $e_A^N=z \delta_A^N$ is the inverse vielbein, $D_N=\partial_N+\frac{1}{8} \omega_{N A B}\left[\Gamma^A, \Gamma^B\right]-i V_N$ is the covariant derivative which ensures the action satisfies gauge invariance and diffeomorphism invariance, and $M$ is the mass of the bulk spinor. The Dirac gamma matrices are defined in such a way that they satisfy the anticommutative relation $\left\{\Gamma^A, \Gamma^B\right\}=2 \eta^{A B}$. The background dilaton field is given by $\Phi(z)=\kappa^2 z^2$, and the right and left spinor are defined as $\Psi_{R,L}=(1/2)(1\pm\gamma^{5})\Psi$. By imposing appropriate boundary conditions, the normalizable wave function can be expressed as
\begin{equation}
\begin{gathered}
\psi_L^{(n)}(z)=\frac{1}{\kappa^{\alpha-1}} \sqrt{\frac{2 \Gamma(n+1)}{\Gamma(\alpha+n+1)}} \xi^\alpha L_n^{(\alpha)}(\xi), \\
\psi_R^{(n)}(z) =\frac{\sqrt{n+\alpha}}{\kappa^{\alpha-1}} \sqrt{\frac{2 \Gamma(n+1)}{\Gamma(\alpha+n+1)}} \xi^{\alpha-(1 / 2)} L_n^{(\alpha-1)}(\xi),
\end{gathered}
\end{equation}
where $\alpha=M+\frac{1}{2}$ and $\kappa=0.35$ GeV. The correct large momentum scale for the proton electromagnetic form factor is given when $M=\frac{3}{2}$. The present investigation solely focuses on the ground state of the proton.

The electromagnetic current matrix element can be generally expressed in terms of two independent
 form factors,
\begin{equation}
\begin{aligned}
\langle p_{3},s_{3}|J^{\mu}(0)|p_{1},s_{1}\rangle& =u(p_{3},s_{3}){\Big(}f_{1}(Q)\gamma^{\mu}  \\
&+f_2(Q)\frac{i\sigma^{\mu\nu}q_\nu}{2m_n}\bigg)u(p_1,s_1),
\end{aligned}
\end{equation}
where $q=p_3-p_1$ and $Q^2=-q^2$.
The invariant functions are given by
\begin{equation}
    C_1(Q)=\int dze^{-\Phi}\dfrac{V(Q,z)}{2z^{2M}}(\psi_L{}^2(z)+\psi_R{}^2(z)) ,
\end{equation}
\begin{equation}
    C_2(Q)=\int dze^{-\Phi}\frac{\partial_z V(Q,z)}{2z^{2M-1}}(\psi_L{}^2(z)-\psi_R{}^2(z)) ,
\end{equation}
\begin{equation}
    C_3(Q)=\int dze^{-\Phi}\frac{2m_n V(Q,z)}{z^{2M-1}}\:\psi_L(z)\psi_R(z) .
\end{equation}

For the soft-wall model, the bulk-to-boundary propagator of the vector field is written as \cite{factormodel}
\begin{equation}
\begin{aligned}V(Q,z)&=\Gamma(1+a)U(a,0;\xi)\\ &=a\int_0^1dxx^{a-1}\exp\left(-\dfrac{x}{1-x}\xi\right),
\end{aligned}
\end{equation}
where $a=Q^2/(4\kappa^2)$, and $\xi=\kappa^2z^2$.

The electric and magnetic form factor for the proton can be obtained by
\begin{equation}
    G_E(Q)=C_1(Q)+\eta_pC_2(Q)-\tau\eta_{p}C_{3}(Q),\label{teff}
\end{equation}
\begin{equation}
    G_M(Q)=C_1(Q)+\eta_{p}C_2(Q)+\eta_{p}C_{3}(Q),
\end{equation}
where $\eta_{p}=0.224,\tau=\frac{Q^2}{4 m_p^2}$. And the effect of both form factors can be described by the effective electromagnetic form factor squared,
\begin{equation}
G_{e f f}^2(t)=\frac{1}{1+\tau}\left[G_{\mathrm{E}}^2(t)+\tau G_{\mathrm{M}}^2(t)\right].
\end{equation}

\subsection{Coulomb interaction amplitude}
In studies of $pp$ and $p\Bar{p}$ scattering, the Coulomb interaction becomes exceedingly prominent as the momentum transfer $t$ approaches zero. Typically, the standard way to represent the whole scattering amplitude was expressible in the form~\cite{Bethe:1958zz}
\begin{equation}
F_{tot}=F_N+e^{i \alpha \phi} F_C.\label{tot}
\end{equation}

The lowest-order one-photon-exchange Coulomb amplitude for pointlike charges is
\begin{equation}
   F_C(s,t)=\mp\dfrac{8\pi\alpha s}{|t|},
\end{equation}
where $\alpha$ is the fine structure constant. The negative (positive) sign corresponds to the scattering of particles possessing identical (opposing) charges.

In Eq.~\eqref{tot}, the amplitudes of strong interaction and Coulomb interaction are bound mutually with the help of the additional phase difference $\alpha \phi(s, t)$ which was the result of the possibility of multiphoton exchange processes.

The Coulomb phase $\phi$  was calculated first by Bethe with the WKB approach in potential theory and derived the following form~\cite{Bethe:1958zz}:
\begin{equation}
\phi=2 \ln \left(1.06 /\left|\mathbf{k}_{\mathbf{1}}\right| b \Theta\right),
\end{equation}
where $\left|\mathbf{k}_1\right|$ is the c.m. momentum, $b$ is the range of the strong-interaction forces, and $\Theta$ is the c.m. scattering angle. Similar results were obtained by these authors \cite{simila1,simila2} using the potential model

West and Yennie (WY) re-examined the interference between Coulomb interaction and strong interaction in terms of Feynman diagrams~\cite{WYphase}
\begin{equation}
\phi_{W-Y}=\mp[\ln (B|t| / 2)+\gamma+O(B|t|)],
\end{equation}
where $\gamma$ is the Euler constant. The upper (lower) sign corresponds to the scattering of $p p$ $(\bar{p} p)$. $B$ is the $t$-independent diffractive slope of the strong-interaction amplitude and is associated with the center of mass energy $\sqrt{s}$, generally defined as
\begin{equation}
B(s, t)=\lim _{t \rightarrow 0} \frac{d\left[\ln \left(d \sigma_N / d t\right)\right]}{d t}.
\end{equation}

R. Cahn~\cite{Cahn} gives a more precise calculation based on the above which accounts for the details of the electromagnetic form factor under the assumption that $|t|\rightarrow0$, and obtained a general expression for the phase
\begin{equation}
\begin{aligned}
\phi_{C a h n}= -[\ln \left(\frac{B|t|}{2}\right)+\gamma+C],\label{phase}
\end{aligned}
\end{equation}
\begin{equation}
    C=\ln \left(1+\frac{8}{B \Lambda^2}\right)+\left(4|t| / \Lambda^2\right) \ln \left(4|t| / \Lambda^2\right)+2|t| / \Lambda^2,
\end{equation}
where $\Lambda^2=0.71$ $\rm{GeV^2}$. In the present work, we will determine $\Lambda^2$ by comparing the data with the electromagnetic form factor which we previously introduced.

The main difference from the result of WY is a shift of the Coulomb amplitude due to the form factor¡¯s influence on the phase. Compared with WY, the Coulomb phase calculated with Cahn's showed a better fit to the experimental data~\cite{Petrov}. Furthermore, as noted in Ref.~\cite{Donnachie:2022aiq},  the form of the Coulomb phase proposed by WY contradicts the general properties of analyticity in the $t$ channel. And from Ref.~\cite{Prochazka:2017tby}, we can know that the theoretical assumptions of the WY model were inconsistent with experimental data. In addition, Nurushev~\cite{Nurushev} and Kopeliovich~\cite{Kopeliovich} derived the Coulomb phase in a large range of momentum transfer, and the results were similar to the calculation of Cahn. Considering the kinematic range of this work, we choose Cahn's calculation for the Coulomb phase.

\begin{figure}[h]
\centering
\includegraphics[scale=0.6]{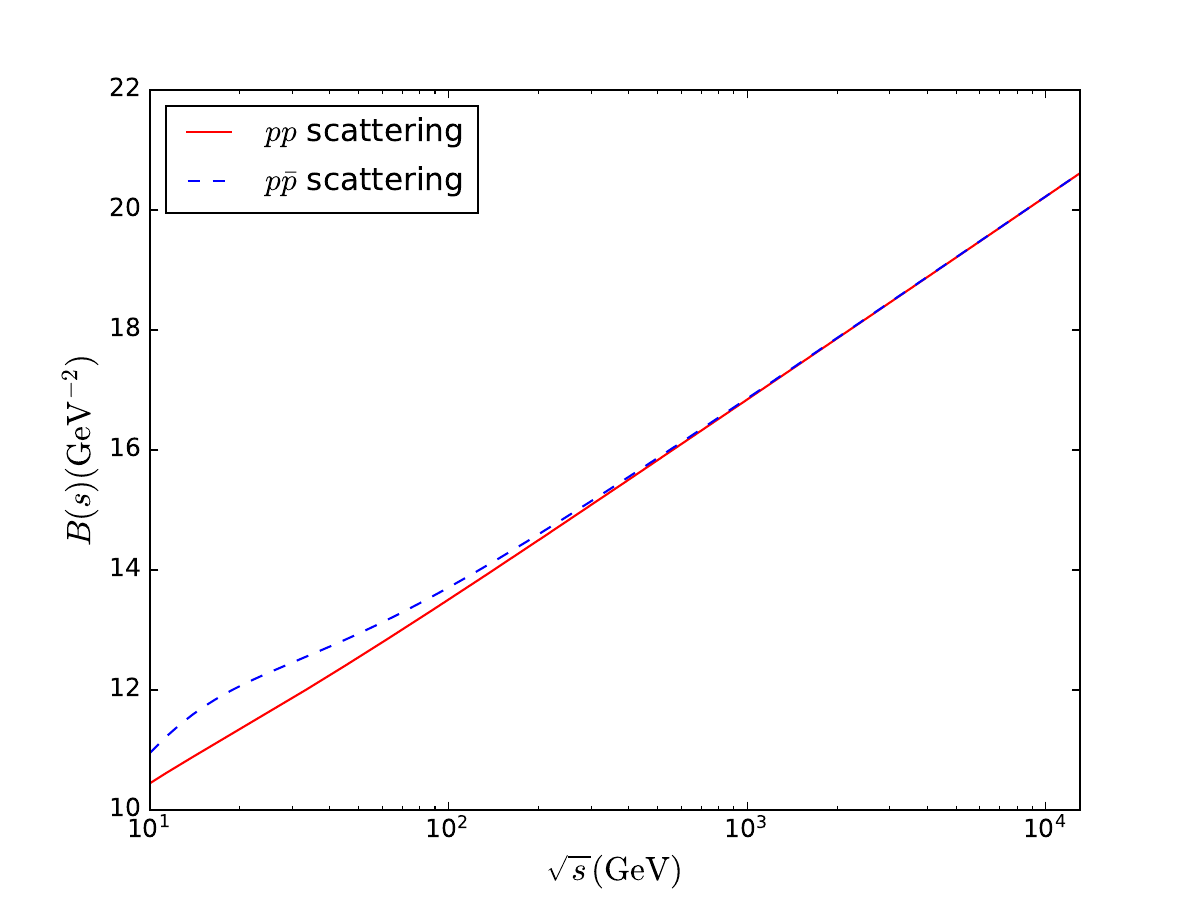}
\caption{The $t$ slope of the strong-interaction amplitude as a function of $\sqrt{s}$ for the $p p$ and $p \Bar{p}$ scattering.}
\label{B}
\end{figure}

By relating to the strong-interaction amplitude, we give the trend of $t$-slope $B$ with the center of mass energy $\sqrt{s}$ for $pp$ and $p\Bar{p}$ scattering, as shown in Fig.~\ref{B}. It can be observed that the $t$ slope $B$ exhibits a consistent increasing trend, and the difference in $t$ slope values between $pp$ scattering and $p\Bar{p}$ scattering is negligible when the energy $\sqrt{s}$ is greater than 100 GeV. According to Ref.~\cite{liu2023pomeron}, it is evident that the contribution of Reggeon and its cross term with Pomeron to the differential cross section in strong interactions can be disregarded when the energy $\sqrt{s}$ is greater than 100 GeV. When the energy $\sqrt{s}$ is below 100 GeV, there are some differences in the B values of $pp$ and $p\Bar{p}$ scattering due to the varying parameters involved in the Reggeon exchange.

Following the introduction of the proton electromagnetic form factor, the Coulomb interaction amplitude can be written as
\begin{equation}
F_C(s, t)=-\frac{8 \pi \alpha s}{|t|} G_{e f f}^2(t).
\end{equation}

Then we obtain the total scattering amplitude that includes both the Coulomb and strong interaction.
\begin{equation}
\begin{aligned}
 F_{tot}=& -e^{i \alpha \phi} \frac{8 \pi \alpha s}{|t|} G_{e f f}^2(t)-s \lambda_{g}^{2} A^{2}(t) e^{-\frac{i \pi \alpha_{g}(t)}{2}} \frac{\Gamma\left[-\chi_{g}\right] \Gamma\left[1-\frac{\alpha_{g}(t)}{2}\right]}{\Gamma\left[\frac{\alpha_{g}(t)}{2}-1-\chi_{g}\right]}\left(\frac{\alpha_{g}^{\prime} s}{2}\right)^{\alpha_{g}(t)-1} \\
& +s \lambda_{v}^{2} \alpha_{v}^{\prime} e^{-\frac{i \pi \alpha_{v}(t)}{2}} \sin \left[\frac{\pi \alpha_{v}(t)}{2}\right]\left(\alpha_{v}^{\prime} s\right)^{\alpha_{v}(t)-1} \Gamma\left[-\alpha_{v}(t)\right],
\end{aligned}
\end{equation}
and the total differential cross section is given by
\begin{equation}
\frac{d \sigma_{tot}}{d t}=\frac{1}{16 \pi s^2}\left| F_{tot}(s, t)\right|^2.
\end{equation}

\section{Numerical results}\label{3}
\subsection{Contribution ratios of the Coulomb and strong interaction}
\begin{figure}[htbp]
\centering
\includegraphics[scale=0.33]{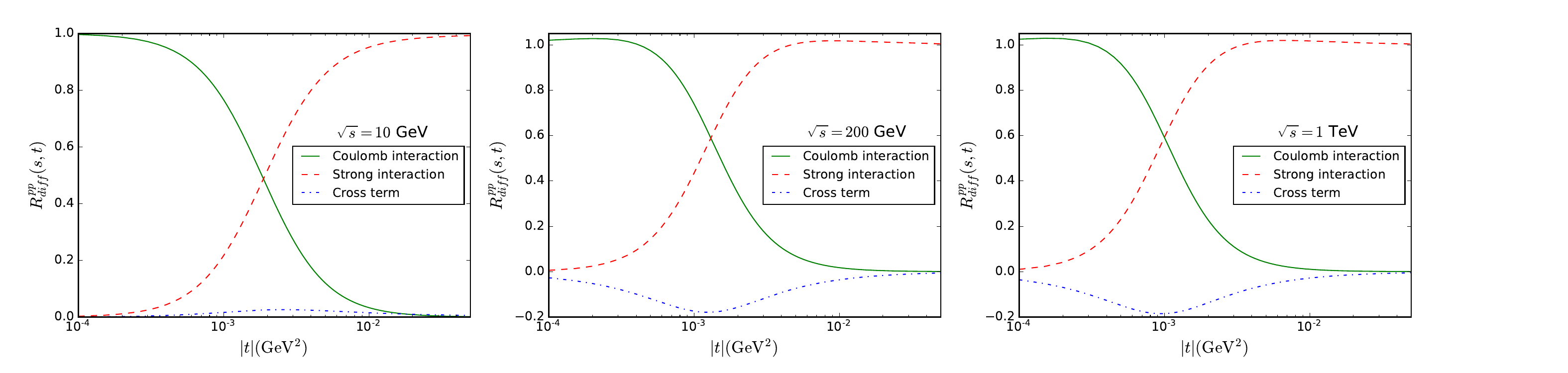}
\caption{The contribution ratios for the differential cross section as a function of $|t|$ for $pp$ scattering. The solid, dashed, and dash-dot curves represent results for the Coulomb interaction, the strong interaction, and the cross term, respectively.}
\label{R}
\end{figure}

\begin{figure}[htbp]
\centering
\includegraphics[scale=0.33]{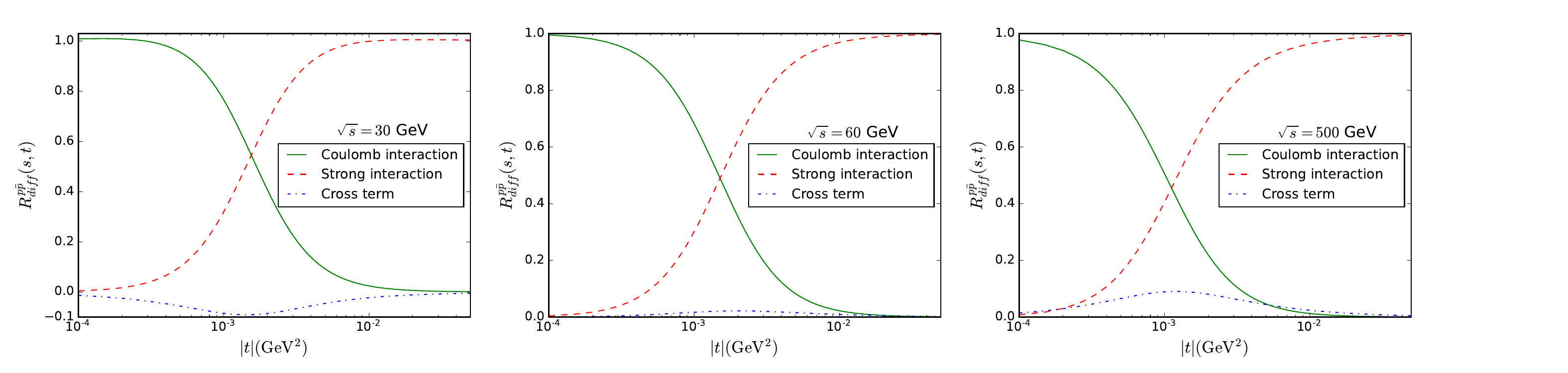}
\caption{The contribution ratios for the differential cross section as a function of $|t|$ for $p\Bar{p}$ scattering. The solid, dashed, and dash-dot curves represent results for the Coulomb interaction, the strong interaction, and the cross term, respectively.}
\label{R1}
\end{figure}

Here we present the $t$ dependence of contributions of the strong interaction,  Coulomb interaction, and the cross term in the present model. We numerically evaluate the contribution of these three items to the total differential cross section for the $pp$ and $p\Bar{p}$ scattering, respectively. Considering the applicability of our present model and taking into account the range of  Coulomb interaction, we decide to focus on the kinematic region, where $10 \enspace\rm{G e V}\leq\sqrt{s} \leq 13 \enspace\rm{T e V}$ and $0 \leq|t| \leq 0.01 \enspace\rm{G e V^2}$. Focusing on these kinematic ranges, we display the $t$ dependence of the ratios for $pp$ scattering in Fig.~\ref{R}, and the ratios for $p\Bar{p}$ scattering in Fig.~\ref{R1}. The Coulomb interaction contribution decreases with $|t|$, and it is opposite for the strong-interaction contribution. These trends presented are consistent with the previous research~\cite{in45,in46}. When the momentum transfer $t$ is very small, the impact parameter $b$ in scattering becomes larger, and the electromagnetic interaction dominates. As the momentum transfer $t$ gradually increases, the impact parameter $b$ continuously decreases, and thus the strong interaction begins to dominate. Owing to the electric charge of the proton, the contribution of the cross term occasionally exhibits a negative value. Regardless of whether the contribution of the cross term is positive or negative, with the variation of t, there will be either a maximum or minimum value when the contribution of the Coulomb and strong interaction are approximately equal. In the case of $pp$ scattering, as the energy $\sqrt{s}$ increases, the contribution of the cross term undergoes a transition from positive to negative values while continuously decreasing. For $p\Bar{p}$ scattering, the trend is precisely opposite to $pp$ scattering. As the energy $\sqrt{s}$ increases, the contribution of the cross term transitions from negative values to gradually become positive values, while continuously increasing.

\subsection{Differential cross section}
In the preceding section, we obtained the parametrized form of the differential cross section, derived from the characterization of the total scattering amplitude.
In the strong-interaction section, seven adjustable parameters are given in the expression for the invariant amplitude, and there is no additional adjustable parameter in the form factors and Coulomb interaction amplitude. For these seven parameters, as previously stated, we use the values obtained in the previous work~\cite{liu2023pomeron}.

By calculating the contribution to the Coulomb amplitude, the Coulomb contribution and the cross term contribution are almost nonexistent at $|t| = 0.05$, as shown in Fig.~\ref{R}; we use the $\Lambda^2$ obtained by matching with the electromagnetic form factor $G_{e f f}^2(t)$  in the range of $0 \leq|t| \leq 0.05\enspace\rm{G e V^2}$. By utilizing the Scipy package of Python, one obtains, for both the $pp$ and $p\Bar{p}$ scattering, $\Lambda^2=0.69$ $\rm{GeV^2}$.

\begin{figure}[h]
\centering
\includegraphics[scale=0.32]{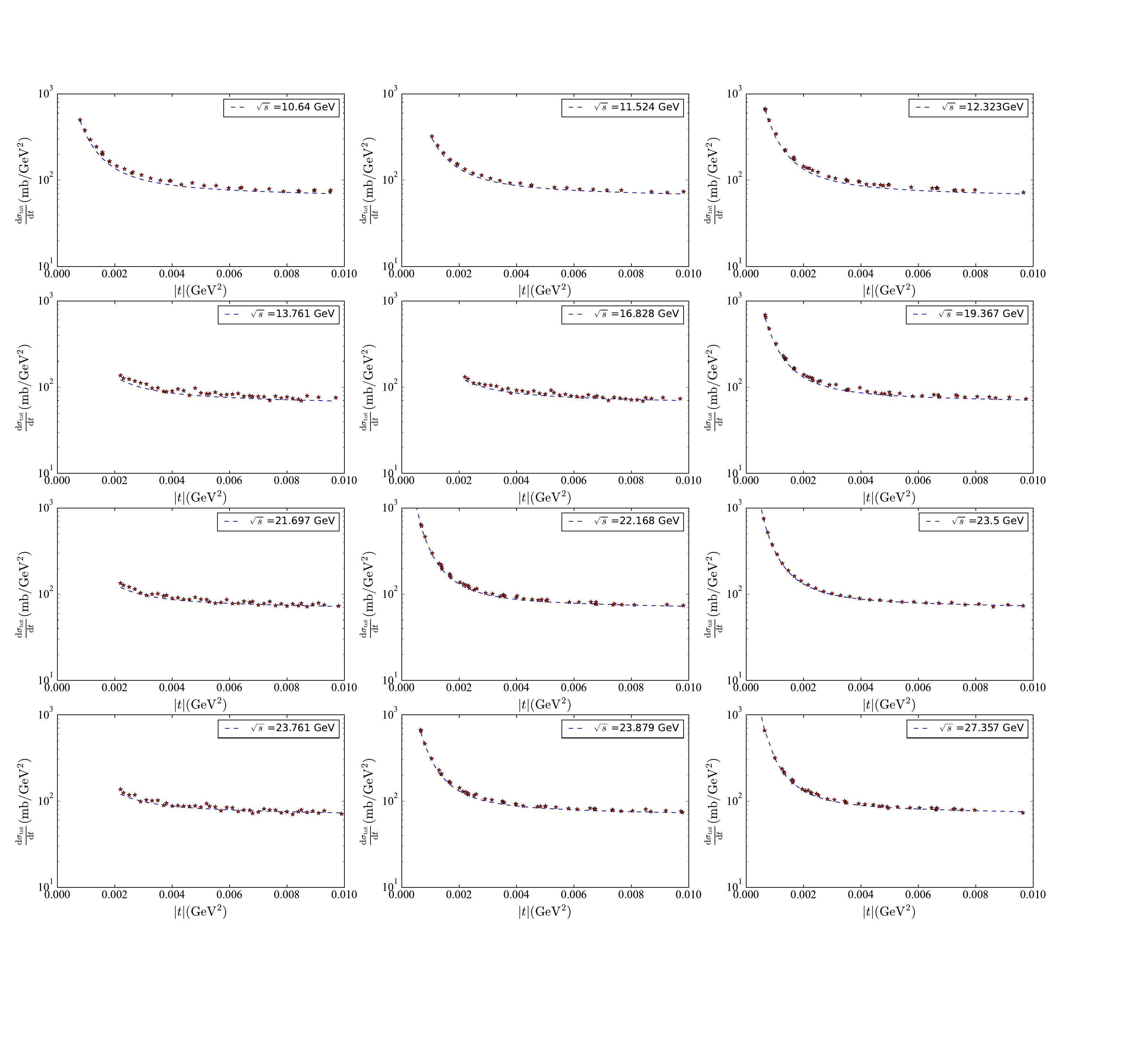}
\caption{The total differential cross section of the $p p$ scattering as a function of $|t|$ for $5\ \text{GeV}<\sqrt{s}<30\ \text{GeV}$. The dashed curves represent the results of our calculations, and the experimental data are expressed as stars.}
\label{dt1}
\end{figure}

\begin{figure}[h]
\centering
\includegraphics[scale=0.32]{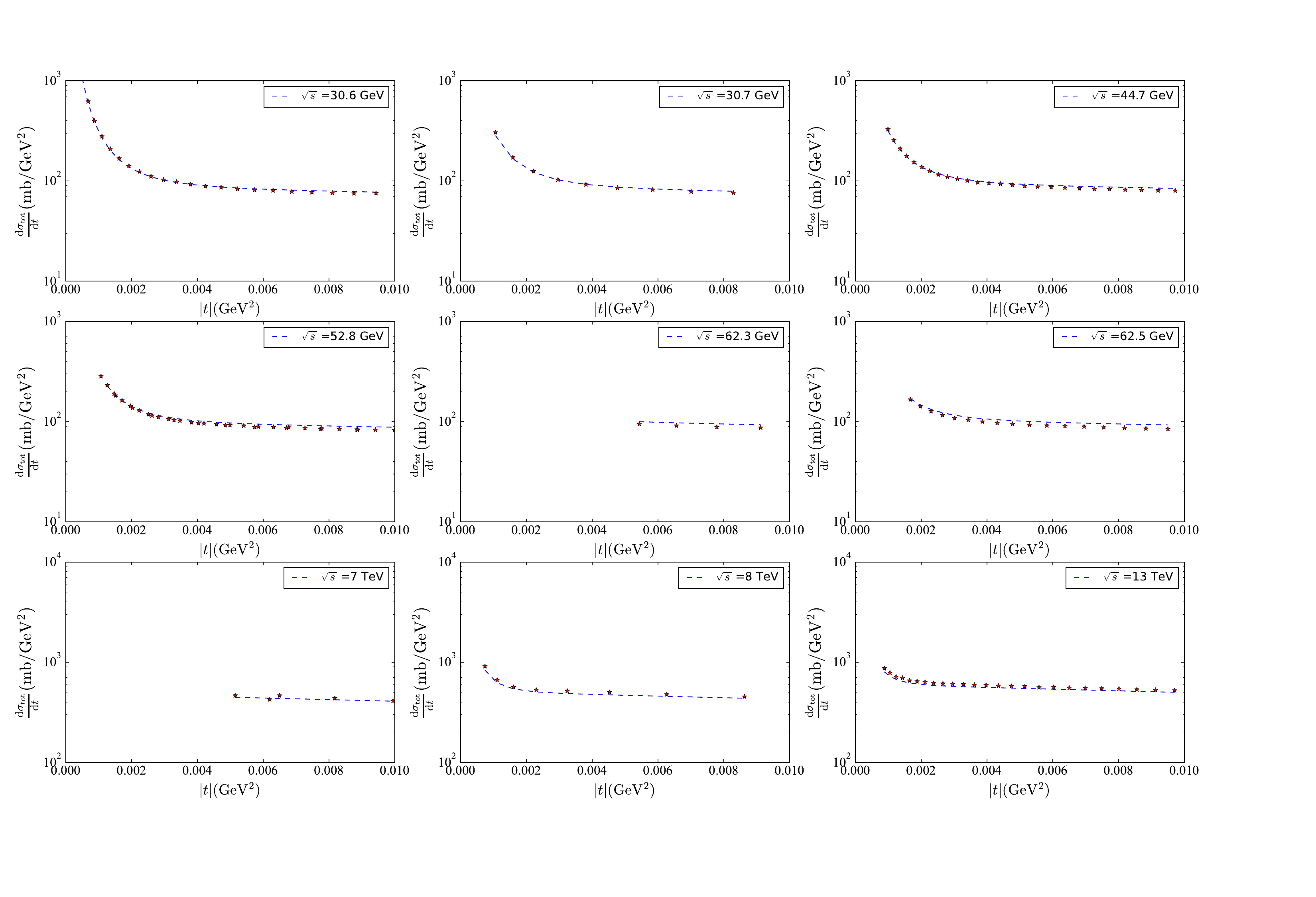}
\caption{The total differential cross section of the $p p$ scattering as a function of $|t|$ for $30\ \text{GeV}<\sqrt{s}<13\ \text{TeV}$. The dashed curves represent results for our calculations, and the experimental data are expressed as stars.}
\label{dt2}
\end{figure}

\begin{figure}[h]
\centering
\includegraphics[scale=0.32]{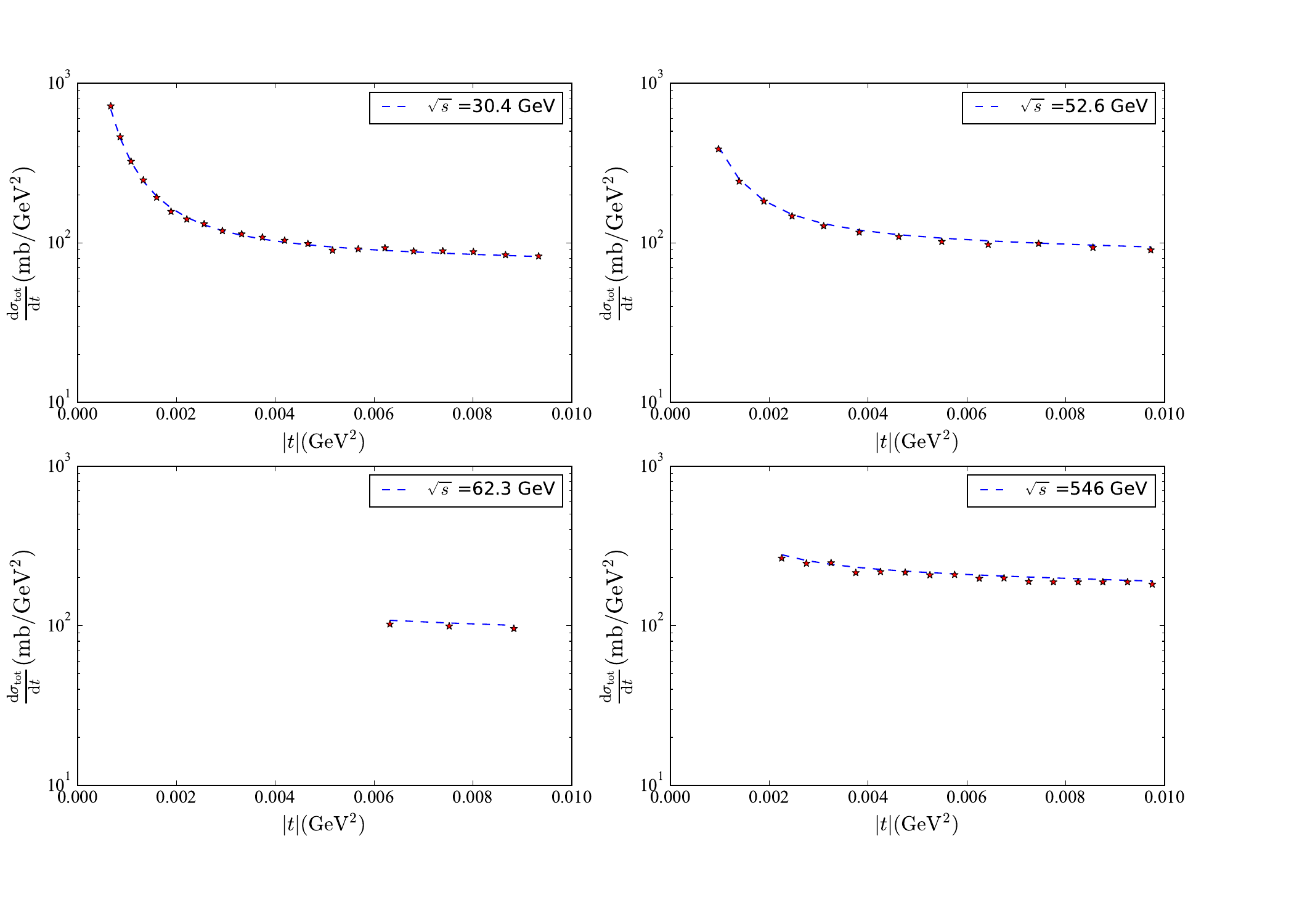}
\caption{The total differential cross section of the $p \Bar{p}$ scattering as a function of $|t|$ for $\sqrt{s}>30\ \text{GeV}$. The dashed curves represent results for our calculations, and the experimental data are expressed as stars.}
\label{dt3}
\end{figure}

We present our results of the differential cross section for $pp$ and $p\Bar{p}$ scattering to demonstrate the correction of scattering by Coulomb interaction, focusing on the Regge regime. In the kinematic range being considered, the Coulomb interactions cannot be neglected in the extremely small range of $|t|$, and the cross term between the two interactions also has a crucial role in the model. By combining Coulomb with strong interaction, we plot the differential cross section with fixed center-of-mass energy $\sqrt{s}$. The experimental data that we are using for the $pp$ scattering are taken from Refs.~\cite{pp1,pppb1,pp2,pp3,pp4,pp5,pp6,pp7,pp8}. The results of $pp$ scattering for the kinematic range of $10\ \text{GeV}<\sqrt{s}<30\ \text{GeV}$ are shown in Fig.~\ref{dt1}. Based on the analysis of the data, it is evident that our calculations are in alignment with the overarching trends. The results of $p p$ scattering for $30\ \text{GeV}<\sqrt{s}<13\ \text{TeV}$ are displayed in Fig.~\ref{dt2}. This figure presented herein encompasses a substantial range for $\sqrt{s}$. Notably, our model provides an accurate description of the corresponding data. The experimental data we are using for the $p \Bar{p}$ scattering are taken from Refs.~\cite{pb2,pppb1}. The quantity of available data for $p \Bar{p}$ scattering is notably lower when compared to that of $p p$ scattering. Specifically, at the GeV scale, our collection of datasets for  $p \Bar{p}$ scattering consists of a mere four instances, and no data have been gathered at the TeV scale. The results are presented in Fig.~\ref{dt3}. Although the amount of data collected for $p\Bar{p}$ scattering is limited and may lead to some loss of credibility, it is undeniable that we have still obtained an excellent consistency between our calculated results and the experimental data.

\section{Conclusion}\label{4}
We have investigated the differential cross sections of $p p$ and $p \Bar{p}$ scattering with the incorporation of Coulomb interaction, within the framework of holographic QCD. In our model setup, the strong interaction are considered to be represented by the exchange of Pomeron and Reggeon in the Regge regime. The Pomeron and Reggeon exchanges are described by the Reggeized spin-2 glueball and vector meson propagators, respectively. By combining the proton-vector meson and proton-glueball couplings with those propagators, we have obtained the strong-interaction amplitude. The Coulomb interaction amplitude is represented by the lowest-order pointlike charge amplitude and takes into account the influence of the electromagnetic form factor. According to Bethe~\cite{Bethe:1958zz}, the complete amplitude of a scattering process is typically expressed as a superposition of two distinct contributions -- the Coulomb interaction amplitude and the strong-interaction amplitude. These two amplitudes are connected by the coulomb phase factor that serves to mutually bind them together. We have adopted the classic Coulomb phase Eq.~\eqref{phase}, which has taken into account the form factor of proton.

There are several parameters in our model, but all of which can be fixed with the values obtained in previous works~\cite{liu2023pomeron}. Apart from these parameters, the introduction of  form factors and Coulomb interactions in the model will not introduce any additional parameters. As demonstrated in this work, the total scattering amplitude obtained by combining strong interaction and Coulomb interaction can well describe the physical process of $pp$ and $p\Bar{p}$ scattering which reproduces quite well the data in the interference region without any additional parameters. The differential cross section results we provided have been found to be in excellent agreement with experimental data for both $pp$ and $p\Bar{p}$ scattering.

\begin{acknowledgments}
This work is supported by the National Natural Science Foundation of China Grants under No. 12175100, the Natural Science Foundation of Hunan Province of China under Grant No. 2022JJ40344, and the Research Foundation of Education Bureau of Hunan Province, China (Grant No. 21B0402). And we would like to thank Zhibo Liu for his assistance in programming.
\end{acknowledgments}

\bibliography{ref}

\end{document}